\documentclass[aps,10pt,prl,superscriptaddress,preprintnumbers,%
showkeys,nofootinbib,showpacs,twocolumn]{revtex4-1}
\usepackage{bm} 
\usepackage{graphicx}
\usepackage{hyperref}
\newcommand{\be}{\begin{eqnarray}}
\newcommand{\ee}{\end{eqnarray}}
\newcommand{\beq}{\begin{equation}}
\newcommand{\eeq}{\end{equation}}
\begin{document}
\title{Quantum--mechanical picture of peripheral chiral dynamics} 
\author{C.~Granados}
\affiliation{Department of Physics and Astronomy, 
Nuclear Physics, Uppsala University, 75120 Uppsala, Sweden}
\author{C.~Weiss}
\affiliation{Theory Center, Jefferson Lab, Newport News, VA 23606, USA}
\begin{abstract}
The nucleon's peripheral transverse charge and magnetization densities 
are computed in chiral effective field theory. The densities are 
represented in first--quantized form, as overlap integrals of chiral 
light--front wave functions describing the transition of the nucleon to 
soft pion--nucleon intermediate states. The orbital motion of the pion 
causes a large left--right asymmetry in a transversely 
polarized nucleon. The effect attests to the relativistic nature of 
chiral dynamics [pion momenta $k = O(M_\pi)$] and could be observed in 
form factor measurements at low momentum transfer.
\end{abstract}
\preprint{JLAB-THY-15-2016}
\maketitle
The long--distance behavior of strong interactions is governed by the 
spontaneous breaking of chiral symmetry in the microscopic theory
of Quantum Chromodynamics. The associated Goldstone bosons --- the pions ---
are almost massless on the hadronic scale, couple weakly to other
hadrons (proportional to their momentum), and mediate long--distance
interactions. The resulting ``chiral dynamics'' can be studied
systematically using methods of effective field theory 
(EFT) \cite{Gasser:1983yg,Weinberg:1990rz}
and explains numerous phenomena in low--energy pion--pion
and pion--nucleon scattering, the nucleon--nucleon interaction at large 
distances, and electroweak interactions of hadrons.

Chiral dynamics represents an essentially relativistic dynamical system, 
as the pion 4--momenta in typical processes are of the order of the 
pion mass, $k = O(M_\pi)$ \cite{Gasser:1983yg}, 
and the number of particles changes 
due to quantum fluctuations. Chiral EFT is therefore usually
formulated and solved as a second--quantized field theory. While this
allows one to calculate most observables of interest, for many
purposes it would be desirable to have a first--quantized, 
particle--based formulation of the dynamics. It would make it
possible to follow the space--time evolution of chiral processes and
gain a more intuitive understanding of their effects. It would
introduce the concept of a wave function and its densities and
help quantify the spatial structure of hadrons, the orbital 
motion of pions, and polarization effects.

The light--front (LF) formulation of relativistic 
dynamics \cite{Dirac:1949cp,Leutwyler:1977vy,Brodsky:1997de} makes 
it possible to construct a consistent first--quantized description of 
essentially relativistic systems.
In this framework one follows the evolution of 
the system in LF time $x^+ = x^0 + x^3 \equiv t + z$.
The wave functions at fixed $x^+$ are invariant under Lorentz boosts
in the longitudinal ($z-$) direction, so that their particle content and
densities are frame--independent and have objective meaning --- in contrast
to the equal--time wave function, where they are frame--dependent.
Transverse boosts (in the $x, y$--plane) are kinematical and preserve the
particle number. Orbital motion and spin are naturally expressed
and lead to a description in close correspondence to non-relativistic
quantum mechanics \cite{Brodsky:1997de}.

In this work we use chiral EFT in the LF formulation to compute the
long--distance contributions to the nucleon's electromagnetic
current matrix element and explain their properties.
The form factors are expressed in terms of the transverse densities 
of charge and magnetization at fixed LF 
time \cite{Soper:1976jc,Burkardt:2000za,Burkardt:2002hr,Miller:2007uy}. 
We calculate the 
isovector densities at peripheral transverse distances $b = O(M_\pi^{-1})$ 
using chiral EFT in the leading--order (LO) approximation. We represent the 
densities in first--quantized form, as overlap integrals of chiral LF 
wave functions describing the transition of the nucleon to soft 
pion--nucleon intermediate states. The new representation leads to 
a simple quantum--mechanical picture, according to which the orbital 
motion of the soft pion causes a left-right asymmetry of the ``plus''
current density in a transversely polarized 
nucleon \cite{Burkardt:2002hr}. The effect
is sizable and attests to the essentially relativistic nature 
of chiral dynamics. Details will be presented elsewhere \cite{long}.

\begin{figure}[t]
\includegraphics[width=.24\textwidth]{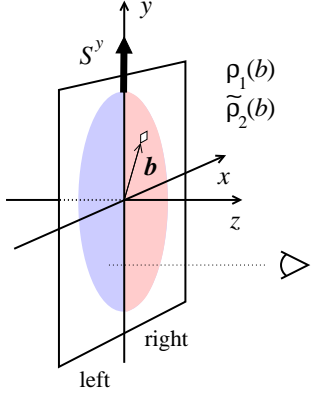}
\caption[]{Interpretation of the transverse densities in a 
nucleon state polarized in the $y$--direction, Eq.~(\ref{j_plus_rho}).}
\label{fig:interpretation}
\end{figure}

The transition matrix element of the electromagnetic current between 
nucleon states with 4--momenta $p_1$ and $p_2$
is parametrized by the Dirac and Pauli form factors, $F_{1}(t)$
and $F_{2}(t)$, which are functions of the invariant momentum transfer 
$t \equiv \Delta^2 = (p_2 - p_1)^2$ (we follow the notation of 
Ref.~\cite{Granados:2013moa}). In a frame where the momentum transfer 
is transverse, $\bm{\Delta}_T \equiv (\Delta^x, \Delta^y) \neq 0, \,
\Delta^{0} = \Delta^z = 0$,
the form factors are represented as a Fourier integral over a
transverse coordinate $\bm{b} \equiv (b^x, b^y)$ 
\cite{Burkardt:2000za,Miller:2007uy}
\beq
F_{1, 2}(t = -\bm{\Delta}_T^2) \;\; = \;\; \int d^2 b \; 
e^{i \bm{\Delta}_T \cdot \bm{b}} \; \rho_{1, 2} (b) .
\label{rho_def}
\eeq
The functions $\rho_{1, 2} (b \equiv |\bm{b}|)$ describe the transverse
spatial distribution of charge and magnetization in the nucleon at fixed 
LF time. Specifically, in a state where the nucleon is localized in 
transverse space at the origin,
and spin--polarized in the $y$--direction, 
the expectation value of the current $J^+ \equiv J^0 + J^3$ 
at LF time $x^+ = 0$ and transverse position $\bm{x}_T = \bm{b}$ is
\be
\langle J^+ (\bm{b}) \rangle_{\text{\scriptsize loc}}
&=& (...) [
\rho_1 (b) \; + \; (2 S^y) \, \cos\phi \, \widetilde\rho_2 (b)] ,
\label{j_plus_rho}
\\[1ex]
\widetilde\rho_2 (b) &\equiv& \frac{\partial}{\partial b} 
\left[ \frac{\rho_2(b)}{2 M_N} \right] ,
\label{rho_2_tilde_def}
\ee
where $(...)$ hides a trivial factor arising from the normalization 
of states, $\cos\phi \equiv b^x/b$, and $S^y = \pm 1/2$ is 
the $y$--spin projection in the nucleon rest frame 
(see Fig.~\ref{fig:interpretation}) \cite{Burkardt:2000za,Granados:2013moa}. 
Thus $\rho_1(b)$ describes the 
spin--independent (left--right symmetric)
and $\cos\phi \, \widetilde\rho_2 (b)$
the spin--dependent (left--right antisymmetric) 
plus current in the $y$--polarized nucleon. 
Choosing $S^y = + 1/2$ and looking at two opposite 
points on the $x$--axis, $\bm{b} = \mp b \bm{e}_x$, one has 
\be
\langle J^+ (\mp b\bm{e}_x) \rangle_{\text{\scriptsize loc}}
&=& (...) \; [ \rho_1 (b) \, \mp \, \widetilde\rho_2 (b) ] 
\nonumber
\\[1ex]
&\equiv& (...) \; \rho_{\rm left/right}(b),
\label{j_left_right}
\ee
which shows that $\rho_1(b)$ and $\widetilde\rho_2(b)$ can be determined
directly as the left--right symmetric and antisymmetric parts of the plus 
current on the $x$--axis.
%
%

\begin{figure}[t]
\includegraphics[width=.33\textwidth]{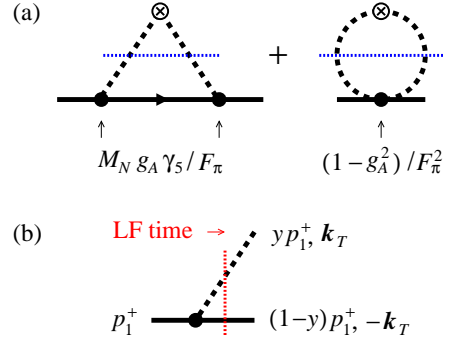}
\caption[]{(a)~Feynman diagrams of LO chiral EFT processes contributing 
to the peripheral transverse densities (two--pion cut of the form factors). 
Indicated below are the 
effective vertices obtained after isolating the nucleon pole term 
of the first diagram. (b)~Chiral LF wave function of the nucleon.}
\label{fig:diag}
\end{figure}

At peripheral distances $b = O(M_\pi^{-1})$ the transverse densities 
are governed by chiral dynamics and can be computed from first
principles using chiral EFT \cite{Granados:2013moa,Strikman:2010pu}. 
The densities can be obtained from the relativistic chiral EFT 
results for the form 
factors \cite{Gasser:1987rb,Bernard:1992qa,Kubis:2000zd}. 
Peripheral contributions arise from
the chiral processes in which the current couples to the nucleon 
through two--pion exchange, i.e., contributions to the two--pion cut 
of the isovector form factors at $t > 4 M_\pi^2$. At LO 
these are given by the Feynman diagrams of Fig.~\ref{fig:diag}a, 
where the vertices are those of the relativistic chiral 
Lagrangian \cite{Becher:1999he}. The first diagram
contains a term in which the pole of the intermediate nucleon 
propagator is canceled
by the numerator; this term is of the same form as the contact term 
from the second diagram and can be combined with it. Effectively this 
amounts to replacing the $\pi NN$ vertices in the first diagram by the 
pseudoscalar vertex $M_N g_A \gamma_5/F_\pi$, and changing the 
$\pi\pi NN$ contact coupling in the second 
as $1/F_\pi^2 \rightarrow (1 - g_A^2)/F_\pi^2$
\cite{Strikman:2010pu,Granados:2013moa}. 
%
%
With this rearrangement the first Feynman diagram is given entirely by 
the nucleon pole contribution. It can therefore be represented as a LF 
time--ordered process in which the initial nucleon makes a transition 
to a soft pion--nucleon intermediate state and back \cite{long}. 
The transition is described by the chiral LF wave function 
(Fig.~\ref{fig:diag}b)
\be
\Psi (y, \bm{k}_T; \, \textrm{pol})
&\equiv& 
\frac{\Gamma (y, \bm{k}_T; \, \textrm{pol})}
{\Delta\mathcal{M}^2 (y, \bm{k}_T)} ,
\label{psi_restframe}
\ee
where $y = k^+/p_1^+$ is the LF plus momentum fraction of the pion,
$\bm{k}_T$ its transverse momentum relative to the initial nucleon
with $\bm{p}_{1T} = 0$, and ``pol'' denotes generic spin quantum
numbers characterizing the initial and intermediate nucleon. In the numerator 
$\Gamma$ is the on-shell pseudoscalar $\pi NN$ vertex between 
the initial nucleon and the intermediate one with LF momentum 
$(1 - y) p_1^+$ and $-\bm{k}_T$.
In the denominator $\Delta\mathcal{M}^2$ denotes the invariant
mass difference between the initial and intermediate state,
\be
\Delta\mathcal{M}^2 (y, \bm{k}_T)
&\equiv& [\bm{k}_T^2 + M_T^2(y)]/[y(1-y)],
\label{invariant_mass_restframe}
\\[1ex]
M_T^2(y) &\equiv& (1 - y) M_\pi^2 +y^2 M_N^2 ,
\label{M_T_def}
\ee
which is proportional to the LF energy denominator of the 
transition \cite{Brodsky:1997de}. The wave function for a state moving with
overall transverse momentum $\bm{p}_{1T} \neq 0$ is obtained by 
a transverse boost, and analogous formulas describe 
the transition back to the final state with $\bm{p}_{2T}$. 
The chiral wave functions
refer to the parametric regime $|\bm{k}_T| = O(M_\pi)$ and 
$y = O(M_\pi/M_N)$, where the pion is soft and couples weakly
to the nucleon, and are used in this context only.
The coordinate--space wave function is
\be
\Phi(y, \bm{r}_T, \, \textrm{pol}) &\equiv&
\int \frac{d^2 k_{T}}{(2\pi)^2} \; e^{i \bm{k}_T \cdot \bm{r}_T} 
\; \Psi (y, \bm{k}_T; \, \textrm{pol}) ,
\label{psi_coordinate}
\ee
where $\bm{r}_T$ is the transverse separation of the pion--nucleon
system in the intermediate state and $|\bm{r}_T| = O(M_\pi^{-1})$.

The peripheral transverse densities can be expressed as overlap integrals 
of the chiral LF wave functions of the initial and final nucleon and an
effective contact term \cite{long}. A particularly 
simple form is obtained when the nucleon spin states in the wave function 
are quantized in the transverse $y$--direction.
Transversely polarized LF spinors are constructed by 
preparing a transverse spinor in the nucleon rest frame and performing
a longitudinal and a transverse boost to get to the desired LF 
momentum \cite{Brodsky:1997de}. We denote the LF wave function 
Eq.~(\ref{psi_coordinate}) definied with such transversely polarized
nucleon spin states by
\beq
\Phi_{\rm tr} (y, \bm{r}_T; \tau, \tau_1), 
\eeq
where $\tau_1$ and $\tau$ are the $y$--spin quantum numbers of the 
initial and the intermediate nucleon states; the complex conjugate
function $\Phi^\ast_{\rm tr} (y, \bm{r}_T; \tau, \tau_2)$ describes the
transition back to the final state with $y$--spin $\tau_2$. 
At the special points 
$\bm{b} = \mp b \bm{e}_x$ [cf.~Eq.~(\ref{j_left_right})]
only the transverse spin--flip wave function ($\tau_1 = \tau_2 = +1/2,
\tau = -1/2$) contributes to the current matrix element;
the spin--nonflip wave function ($\tau = +1/2$) vanishes on the
transverse $x$--axis. We obtain the isovector densities as 
[$\rho^V \equiv (\rho^p - \rho^n)/2$]
\be
\left. 
\begin{array}{l}
\rho_{\rm left}^V (b) \\[.5ex] \rho_{\rm right}^V (b) 
\end{array}
\right\}
&=& 
\int_0^1 \! dy
\, 
\frac{| \Phi_{\rm tr} (y, \mp r_T \bm{e}_x; -\frac{1}{2}, \frac{1}{2}) |^2}
{2\pi y (1 - y)^3} 
\nonumber 
\\[0ex]
&& [r_T = b/(1 - y)] .
\label{rho_overlap_left_right}
\ee
The explicit expressions for the spin--flip wave function 
at large separations $r_T \gg M_T^{-1}$ are [cf.\ Eq.~(\ref{M_T_def})]
\be
\lefteqn{\Phi_{\rm tr} (y, \mp r_T \bm{e}_x; -{\textstyle\frac{1}{2}}, 
{\textstyle\frac{1}{2}})}
\nonumber 
\\[1ex]
&=& \frac{g_A M_N \, y \sqrt{1 - y}}{2 \sqrt{2\pi} F_\pi}
\, \left[ - y M_N \mp M_T(y) \right] \;
\, \frac{e^{-M_T(y) r_T}}{\sqrt{M_T(y) r_T}} ; \hspace{2em}
\label{psi_left_right_explicit}
\ee
exact expression are given in Ref.~\cite{long}.
The charge and magnetization densities are then obtained as
\beq
\left.
\begin{array}{l}
\rho^V_1 (b) \\[.5ex] \widetilde{\rho}_2^V (b) 
\end{array}
\right\}
\;\; = \;\; \frac{1}{2} [ \pm \rho_{\rm left}^V(b) + \rho_{\rm right}^V(b)] .
\label{rho_1_2_from_left_right}
\eeq

The effective contact term in Fig.~\ref{fig:diag} describes the 
instantaneous contributions to the current in LF time (zero modes) 
and has to be added to Eq.~(\ref{rho_overlap_left_right}).
The coupling $\propto (1 - g_A^2)$ shows that this term reflects the
nucleon's internal structure due to non-chiral intermediate 
states \cite{long,Granados:2013moa}. Its contribution to the density
is left--right symmetric and amounts to $<$10\% of $\rho_1^V (b)$ 
at $b > 1\, M_\pi^{-1}$. The peripheral densities are thus practically
determined by the wave function overlap Eq.~(\ref{rho_overlap_left_right}).

%
%
\begin{figure}[t]
\includegraphics[width=.48\textwidth]{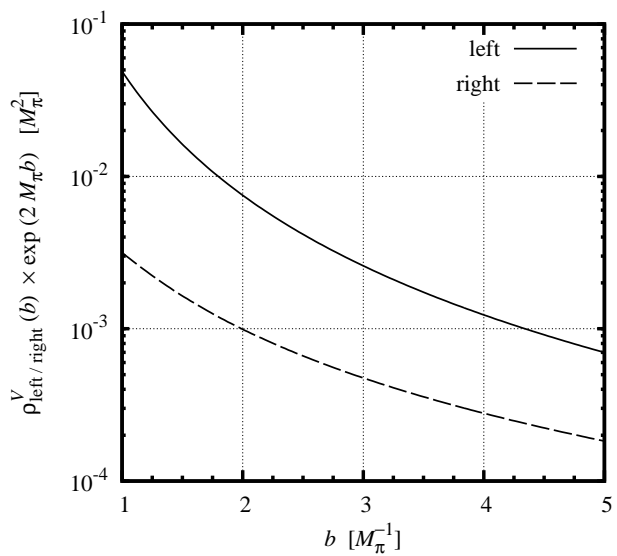}
\caption[]{Left and right peripheral transverse densities
in LO chiral EFT, as given by Eq.~(\ref{rho_overlap_left_right}) and 
the contact term. The plot
shows the densities after extraction of the exponential 
factor $\exp (- 2 M_\pi b)$. The transverse 
distance $b$ and the densities are given in units of the
pion mass.}
\label{fig:rho_left_right}
\end{figure}
The LF representation Eq.~(\ref{rho_overlap_left_right}) (including the 
contact term) is \textit{exactly equivalent} to 
the result of the relativistically invariant EFT 
calculation \cite{Granados:2013moa} and embodies the entire chiral 
structure of the peripheral densities at LO. It reveals 
several interesting properties: (a)~The left and right densities
are of the same parametric order in the heavy--baryon limit, 
$\rho^V_{\rm left}(b)/\rho^V_{\rm right}(b) = O(1)$ for 
$M_\pi / M_N \rightarrow 0$, because the integral in 
Eq.~(\ref{rho_overlap_left_right}) is dominated by pion momentum fractions
$y = O(M_\pi / M_N)$. (b)~The left and right densities in
Eq.~(\ref{rho_overlap_left_right})
are individually positive, $\rho_{\rm left / right}^V(b) > 0$. The
charge and magnetization densities 
Eq.~(\ref{rho_1_2_from_left_right})
therefore obey an inequality,
\beq
|\widetilde\rho_2^V(b)| \; < \; \rho_1^V(b) ,
\label{inequality}
\eeq
as was observed numerically in Ref.~\cite{Granados:2013moa}.
(c)~The left-right asymmetry of the densities produced by chiral
dynamics is numerically large (see Fig.~\ref{fig:rho_left_right}).
The ratio $\rho^V_{\rm left}(b)/\rho^V_{\rm right}(b)$ 
is $\sim$10 at $b = 1 M_\pi^{-1}$ and
decreases slowly at larger distances. As a result the 
charge and magnetization densities Eq.~(\ref{rho_1_2_from_left_right})
are almost equal and opposite,
\beq
\widetilde\rho_2^V(b) \; \approx \; -\rho_1^V(b) ,
\eeq
and the inequality Eq.~(\ref{inequality}) is almost saturated.

\begin{figure}[t]
\includegraphics[width=.48\textwidth]{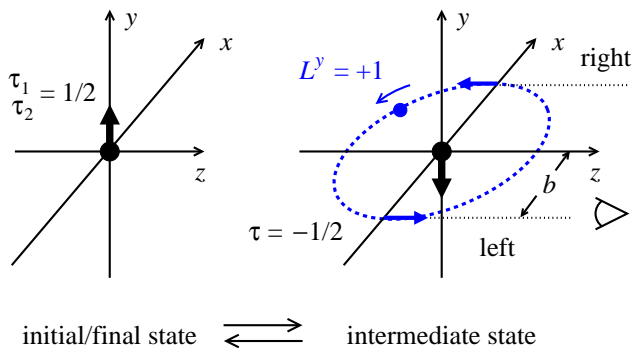}
\caption[]{Quantum--mechanical picture of chiral dynamics in the
peripheral transverse densities (explanation in text).
\label{fig:mech}}
\end{figure}

Our findings can be summarized in a simple quantum--mechanical
picture of the peripheral transverse densities in chiral EFT
(see Fig.~\ref{fig:mech}), inspired by the general arguments of
Ref.~\cite{Burkardt:2002hr}. Consider a physical proton with 
$y$--spin projection $+1/2$ in the rest frame. In the interaction picture 
we may think of this system as a bare nucleon that undergoes 
transitions to multiple pion--nucleon states through the chiral 
EFT interactions. In LO the peripheral left/right densities 
(at the points $\bm{b} = \mp b \bm{e}_x$) arise from the single
$\pi^+ n$ intermediate state, in which the neutron has 
$y$--spin $-1/2$ and the pion is in a state with orbital angular 
momentum $L = 1$ and $y$--projection $L^y = +1$. Because of the
orbital motion the pion on the left moves toward the observer 
and has net positive $z$--momentum $k^z > 0$, while the pion
on the right moves away and has $k^z < 0$. The plus current
carried by a free charged pion is 
$\langle \pi^+(k)| J^+ |\pi^+(k) \rangle = 2 k^+ = 
2(\sqrt{|\bm{k}|^2 + M_\pi^2} + k^z)$.
The observer thus sees a larger plus current on the left than on
the right, resulting in a left--right asymmetry. If the motion of
the pion were non-relativistic, $|\bm{k}| \ll M_\pi$ the asymmetry
would be small, $\rho_{\rm left}/\rho_{\rm right} = 1 + O(|\bm{k}|/M_\pi)$.
That the asymmetry obtained in chiral EFT is large therefore directly 
attests to relativistic motion of the pion, $|\bm{k}| = O(M_\pi)$.

The intuitive arguments presented here assume rotational 
symmetry around the $y$--axis, which is not present in the LF
formulation. The explicit expressions Eqs.~(\ref{psi_left_right_explicit}) 
and (\ref{rho_1_2_from_left_right}) show, however, that all the
described features are realized in the LF formulation as well,
if the nucleon transverse spin states are defined as specified above.
%
%

In sum, the LF formulation of chiral EFT provides a concise representation 
of the peripheral transverse densities, which reveals new properties
(positivity, inequality) and permits a simple mechanical interpretation.
The large left--right asymmetry is rooted in the spin structure of the
pion--nucleon coupling and the essentially relativistic motion of pions
and represents a striking chiral effect.
It could be observed by extracting the peripheral transverse densities
from precise measurements of the nucleon's Dirac and Pauli form factors 
at low momentum transfer, using dispersion fits that respect the
analytic properties \cite{Belushkin:2006qa}; see
Ref.~\cite{Miller:2011du} for details. Similar chiral left--right
asymmetries may be observed in high-energy proton--proton collisions, 
by selecting events in which the scattering takes place on a 
peripheral pion; such processes would permit much more direct tests 
of the effect described here.

The LF wave function representation of peripheral transverse densities
can be extended to include intermediate $\Delta$ isobars and implement
the proper scaling behavior in the large--$N_c$ limit of 
QCD \cite{Strikman:2010pu,Granados:2013moa}. It can also be used
to compute the peripheral densities of matter and angular momentum
(describing the form form factors of the energy--momentum tensor)
and develop a mechanical representation of these structures.
It can be applied further to the nucleon's peripheral parton densities
(generalized parton distributions) \cite{Strikman:2003gz}.
The LF representation has also been employed to study aspects of 
chiral nucleon structure (self--energies, electromagnetic couplings) 
without restriction to peripheral distances \cite{Ji:2009jc}. 

Notice: Authored by Jefferson Science Associates, 
LLC under U.S.\ DOE Contract No.~DE-AC05-06OR23177. The U.S.\ Government 
retains a non--exclusive, paid--up, irrevocable, world--wide license to 
publish or reproduce this manuscript for U.S.\ Government purposes.
\end{document}